\documentclass[11pt,letterpaper,reqno,oneside]{article}\usepackage{booktabs,amsmath,amsfonts,amssymb,graphicx,bm}
\usepackage{bbold}
\usepackage[authoryear]{natbib}
\usepackage{relsize}
\usepackage{times}
\usepackage{dsfont}
\usepackage{subfig}
\usepackage{etoolbox}
\usepackage{float}
\usepackage{epigraph}
\usepackage{ushort}
\usepackage{hyperref}
\usepackage{xcolor}
\usepackage{bbm}
\usepackage{tikz}
\usepackage{setspace}
\usepackage[utf8]{inputenc}
\usepackage[T1]{fontenc}
\usetikzlibrary{arrows,calc, patterns, positioning, shapes.geometric, decorations.pathreplacing,decorations.markings}
\usepackage{relsize}
\usepackage{pgfplots}
\usepgfplotslibrary{fillbetween}
\definecolor{paradise}{RGB}{224, 82, 99}
\usepackage{titlesec}
\titleformat{\section}
{\color{paradise!100}\sffamily\large}
{\color{paradise!100}\thesection.}{.1em}{}
\titleformat{\subsection}[runin] 
{\color{paradise!100}\normalfont\rmfamily\itshape}
{\color{paradise!100}\thesubsection.}{.1em}{}[.]
\titleformat{\subsubsection}[runin] 
{\normalfont\rmfamily\itshape}
{\thesubsubsection.}{.5em}{}[.]

\hypersetup{
  colorlinks   = true, 
  urlcolor     = blue, 
  linkcolor    = blue, 
  citecolor   = blue 
}
\makeatletter
\def\@settitle{\begin{center}%
  \baselineskip14\p@\relax
    \bfseries
    \normalfont\Large\textbf
  \@title
  \end{center}%
}
\usepackage[margin=1in]{geometry}
\usepackage[margin=10pt,font=small,labelfont=bf,
labelsep=period]{caption}

\usepackage{mathtools}

\pgfplotsset{compat=1.18} 
\begin{document}
\renewcommand{\thefootnote}{\fnsymbol{footnote}}

\thispagestyle{empty}
\centerline{\Large \textbf{{\color{paradise}Recent Contributions to Theories of Discrimination}}}

\vspace*{0.2in}

\centerline{Paula Onuchic\footnote[2]{London School of Economics, p.f.onuchic@lse.ac.uk. Many thanks to Daghan Carlos Akkar and Ludvig Sinander for their thorough reading and comments. I'm also grateful for encouragement and helpful feedback from Meg Meyer, Debraj Ray, Peter Hull, Jean-Paul Carvalho, Sam Kapon, Arjada Bardhi, Eduard Krkoska, and Peyton Young.}}

\vspace*{0.2in}

\centerline{December 2025}

\renewcommand*{\thefootnote}{\arabic{footnote}}
\textbf{Abstract.} This paper surveys the literature on theories of discrimination, focusing mainly on new contributions. Recent theories expand on the traditional taste-based and statistical discrimination frameworks by considering specific features of learning and signaling environments, often using novel information- and mechanism-design language;  analyzing learning and decision making by algorithms; and  introducing agents with behavioral biases and misspecified beliefs. An online appendix attempts to narrow the gap between the economic perspective on ``theories of discrimination'' and the broader study of discrimination in the social science literature by identifying a class of models of discriminatory institutions, made up of theories of discriminatory social norms and discriminatory institutional design.

\vspace*{.5in}

\onehalfspacing
Traditionally, economic theories of discrimination are categorized either as taste-based  or as statistical discrimination. The basic taste-based discrimination model in \cite{becker1957} proposes a labor-market environment in which potential employees vary both in terms of their productivity and their identity traits. The model posits that employers' utilities from hiring decisions may depend directly on the identity traits of the workers they choose to employ. A consequence of employers' \emph{bias} for or against members of certain identity groups is that individuals with perhaps the same productivity can be hired at different salaries, or face different probabilities of being hired, based solely on their identities. In this model, a worker's identity is a trait that is payoff-relevant to employers; in contrast, this assumption does not apply to the models of statistical discrimination or to most other models discussed in this survey, which instead see identity as a payoff-irrelevant trait and propose theories to explain identity discrimination despite this payoff-irrelevance. 

Since the taste-based theory proposed by \cite{becker1957} and statistical discrimination theories introduced by \cite{phelps1972} and \cite{arrow1973} (briefly discussed in section \ref{sec:1}), the profession has put much effort towards contrasting these explanations. Much of the enterprise has aimed to propose methods to empirically measure the degree to which differential treatment across identities can be attributed to bias (a term normally used to refer to taste-based discrimination) and/or to a statistical discrimination component. Much of this work is addressed in detail by excellent surveys, such as \cite{arrow1998}, \cite{fangmoro2011}, \cite{langlehmann2012}, \cite{neumark2018}, and \cite{langspitzer2020}, and will not be discussed by this current article.

Instead, this survey article discusses more recent developments in economic theories of discrimination. Section \ref{sec:1} discusses recent generalizations of classic models of statistical discrimination, as well as the adaptation of these theories to the context of algorithmic decision making. Models of statistical discrimination --- including those in section \ref{sec:1} --- typically view learning as described by an exogenous signalling process. Some recent developments, introduced in section \ref{sec:2}, endogenize the decision-maker's learning process. They propose various models of learning environments that support equilibria with discriminatory learning. In section \ref{sec:3}, I review a literature that again considers exogenous learning processes, but allows for processes other than standard Bayesian updating; these include learning with mis-specified models or other behavioral heuristics.

One goal of this review is to highlight that the literature on theories of discrimination has developed closely with advances in economic theory more broadly.\footnote{\cite{bardhiguostrulovici2024} is a contribution that stems from models of strategic experimentation, a la \cite{kellerradycripps2005}; the development of a recent literature on information design leads to theories of discrimination and fairness as in \cite{strackyang2024} and \cite{dovalsmolin2024}; the growing intersection between economic theory and computer science is represented, for example, in \cite{lianglumuokumura2024}; new frameworks in which agents have mispecified models are applied to questions of discrimination, for instance, in \cite{bohrenimasrosenberg2019}.} 
Along with the introduction of new models of discrimination, the recent literature has introduced new proposals on how to regulate markets subject to discrimination: (i) The papers reviewed in section \ref{sec:design4} propose regulatory solutions that are particularly fit to contexts where decisions are made by algorithms, rather than humans. In such environments, a policy-maker has access to different tools, such as regulating the data available to algorithms; (ii) In section \ref{sec:2}, various models introduce the idea that discrimination can result from \emph{learning traps} in the process through which employers gather information about potential employees. Policies that incentivize exploration in this learning process (like the Rooney rule) are particularly effective at remedying these traps; (iii) The behavioral literature that considers learning with mis-specified models or other learning heuristics proposes (and experimentally demonstrates the effectiveness) of policies that target exactly these behavioral biases. 

Recent theoretical developments also mapped into innovations on how to measure and differentiate between different forces behind discrimination, so as to assess the effectiveness of various policies. First, recent generalizations of Phelps' model of statistical discrimination have provided foundations for ``robust'' tests of bias, which do not rely on a particular specification of the learning environment --- see section \ref{sec:gen}. Second, section \ref{sec:inac} proposes methods to identify discrimination stemming from inaccurate beliefs and mis-specified learning environments.


A secondary goal of this review (now allocated to an online appendix) is to discuss earlier literature that is often overlooked by the dichotomical ``bias versus statistical discrimination'' framework. These papers propose theories of discrimination as social norms and show that, in various contexts, the asymmetric treatment of equals can be a feature of optimally designed incentive mechanisms. I interpret these two sections as discussing economic theories of institutional discrimination, taking the view of social norms as informal institutions and incentive mechanisms as formal institutions.

\section{Statistical Discrimination (by Humans or Algorithms)}
\label{sec:1}

\subsection{Phelpsian Statistical Discrimination} 
\label{sec:phelps}

\cite{phelps1972} proposes a simple problem of statistically inferring workers' productivities based on imperfectly informative productivity signals, and workers' \emph{payoff-irrelevant} identities. Each worker has an identity $j\in\{A,B\}$ and  a productivity type $p$ which is drawn from a normal distribution $N(\mu_j,\sigma_j^2)$. An employer wants to hire this worker and is willing to pay a wage equal to the worker's productivity. However, the employer only observes the worker's identity and an imperfect productivity signal, given by $\theta=p+\epsilon$, where $\epsilon$ is distributed according to $N(0,\sigma^2_{\epsilon j})$. 

Given an identity and signal observation, the employer's inference of the worker's expected productivity is
\begin{equation}
\label{eq:1}\mathbb{E}\left(p|\theta,j\right)=\frac{\sigma_j^2}{\sigma_j^2+\sigma^2_{\epsilon j}}\theta+\frac{\sigma_{\epsilon j}^2}{\sigma_j^2+\sigma^2_{\epsilon j}}\mu_j
\end{equation}

If the productivity signal is perfectly informative ($\sigma^2_{\epsilon A}=\sigma^2_{\epsilon B}=0$), then the employer fully observes the worker's type and there is no scope for discrimination. If instead the signal is imperfect, the inferred productivity of a worker may depend on their group identity.

One possible approach is to define discrimination at an individual level: statistical discrimination arises when an employer, upon seeing the same observable outcome (same signal realization $\theta$) from two individuals with different identities, infers different expected productivities, and therefore makes different wage offers to the two individuals.

Equation (\ref{eq:1}) implies that such individual-level discrimination occurs when populations $A$ and $B$ differ, either in their underlying productivity distributions ($\mu_A\neq \mu_B$ or $\sigma^2_A\neq\sigma^2_B$), or in the accuracy of their productivity signal ($\sigma^2_{\epsilon A}\neq\sigma^2_{\epsilon B}$). Suppose $\mu_A<\mu_B$, so that population $A$ is on average less productive than population $B$. Then the inferred productivity of two workers who draw the same signal realization $\theta$ will be different if they belong to different identities, because a worker's identity is effectively an additional signal of their productivity. Specifically, the worker who belongs to $A$ will be expected to be less productive. 

Otherwise, suppose groups have identical productivity distributions, with $\mu_A=\mu_B$ and $\sigma^2_A=\sigma^2_B$, but different productivity signaling technologies. For example, let $\sigma^2_{\epsilon A}>\sigma^2_{\epsilon B}$, so that the signal is more noisy for group $A$ individuals. The productivity inference made by the employer is therefore more responsive to the signal for group $A$ than it is for group $B$. Specifically, a group-$A$ person who draws an above average signal realization is interpreted as having a lower expected productivity than a $B$-group person with the same signal realization. Conversely, a $A$ person with a lower than average signal realization is seen as more productive than if they belonged to group $B$.

In this last exercise, where only the signaling technology differs across groups, both groups receive the same wage on average. To see this, take $\mathbb{E}\left(p|\theta,j\right)$ from equation (\ref{eq:1}), and average it with respect to the distribution of signal realizations $\theta$ generated by each group $j\in\{A,B\}$, to find that $\mathbb{E}\left[\mathbb{E}\left(p|\theta,j\right)\right]=\mu_j$, which is the same across groups. This observation by \cite{aignercain1977} brings into question  definition of discrimination at the individual level in \cite{phelps1972}. \cite{aignercain1977} argue instead that discrimination should be defined at the group level: \emph{Group-level} statistical discrimination arises when two groups with the same underlying productivity distribution, but different productivity-signaling technologies, receive different wages \emph{on average}. 

\cite{aignercain1977} show that group-level discrimination arises if wages offered by employers are a non-linear function of their posterior mean belief about the worker's productivity (perhaps due to the employer being risk-averse). Suppose for instance that  $w:\mathbb{R}\rightarrow\mathbb{R}$ is an increasing and convex wage function, and suppose $\sigma^2_{\epsilon A}>\sigma^2_{\epsilon B}$. Then, using Jensen's inequality, we have $\mathbb{E}\left[w\left(\mathbb{E}\left(p|\theta,A\right)\right)|\sigma^2_{\epsilon A}\right]<\mathbb{E}\left[w\left(\mathbb{E}\left(p|\theta,B\right)\right)|\sigma^2_{\epsilon B}\right]$,
because the distribution of posterior means induced by group $B$'s signal is a mean-preserving spread of the distribution of posterior means induced by group $A$'s signal.\footnote{This argument is not the product of \cite{aignercain1977} model with a risk-averse employer, but rather a simpler illustration of how non-linearity in the wage schedule -- of unmodeled origin -- can yield group level statistical discrimination.} In this case, the group with the more informative signaling technology -- in the sense of Blackwell's garbling order --- receives higher wages on average.

\subsection{Generalizing Phelpsian Statistical Discrimination} \label{sec:gen}
\cite{chambersechenique2021} develop a model of statistical discrimination that generalizes the relationship between group-level discrimination and the signaling technology available to different groups. Their framework differs from Phelps' in two ways: it permits a general structure to the signals seen by employers about prospective employees' underlying types and allows for a general productive technology that determines the mapping between employees' types and their product to their employing firm. In \cite{chambersechenique2021}, a population is defined by an underlying (prior) productivity distribution  and a signaling technology. A signaling technology  is a (perhaps non-deterministic) map between productivity and signal realizations. For example, a signal realization could be a grade at a test, and a signaling technology is the map which specifies a distribution of grades that is attained by people of a given productivity level.

A firm observes a worker's signal realization, as well as their group membership, and forms a posterior about the worker's productivity. The firm has a technology, defined by a set of available tasks and, given their posterior about the worker's productivity, the firm matches the worker with a task, and pays them their productive outcome at the chosen task. This worker-task matching problem performed by the firm yields potentially nonlinear worker-payment schedules. As a variation of \cite{aignercain1977} definition of group-level statistical discrimination, \cite{chambersechenique2021} say statistical discrimination takes place when there is some firm technology such that two populations with the same skill distribution, but different signaling technologies, receive on average different payments. This definition clarifies that the average payments to a population depend not only on the signaling technology, but also on the firm technology. With respect to the latter aspect, it takes a conservative approach in defining discrimination to be present if there is any firm technology that yields an average payment difference.

\cite{chambersechenique2021} show that, if two populations with equal prior productivity distributions have access to signal structures that are ordered according to Blackwell's informativeness criterion, then discrimination ensues, in which the population with the more informative signal is favored. 
Further, they find that discrimination may arise even if signal structures are not ranked according to the Blackwell order: the relevant characterization of Phelpsian statistical discrimination is an identification property. In a follow-up paper, \cite{escudeonuchicsinandervalenzuelastookey2022} observe that identification holds if and only if any two populations with the same skill distribution also have the same signalling technology. A consequence is that discrimination --- at least according to the permissive notion posited by \cite{chambersechenique2021} --- is close to inevitable, only holding when the structure of signals is identical across populations.\footnote{\cite{escudeonuchicsinandervalenzuelastookey2022} show that a version of Blackwell's Theorem, relabeled to fit the statistical discrimination environment, implies a characterization of statistical discrimination that encompasses the results in \cite{chambersechenique2021}. \cite{escudeonuchicsinandervalenzuelastookey2022} is an unpublished working paper, which is subsumed by \cite{escudeetal2025}. This latter paper studies a general statistical discrimination setting, in which employers possibly misperceive the distribution of worker types in each population.}

\cite{bharadwajdebrenou2024} pursue a measurement of bias and statistical discrimination that is based on a generalization of the Phelpsian framework which, like \cite{chambersechenique2021}), allows for a general class of signal structures and wage functions. The perspective in their paper is that of an econometrician who observes realized wage distributions two different populations, and wishes to identify whether differences in these distributions can be attributed to populations having different signaling technologies (statistical discrimination), or must be attributed to employer bias.  This test is ``robust'' precisely because it does not impose a structure to the signalling technologies that may be behind statistical discrimination. \cite{bharadwajdebrenou2024} find that if neither wage distribution dominates the other in the first order stochastic, then differences between the distributions can be ``justified'' by statistical discrimination. If instead one of the wage distributions first-order dominates the other, then the difference in distributions must be at least partly due to employer bias.

\cite{martinmarx2022} also propose a ``robust'' test for bias, in that they characterize situations where differential outcomes across two populations \emph{cannot} be explained by differences in signals. \cite{martinmarx2022}) consider the decision-maker makes a binary decision (to hire or not to hire the agent, say), and the agents' true type is revealed to the econometrician after either decision is made. \cite{martinmarx2022} show that if the average productivity amongst un-hired agents in population $A$ is larger than the average productivity amid hired agents in population $B$, then this is a robust indicator of the decision-maker's bias.

\subsection{Equilibrium Statistical Discrimination} 
\label{sec:eqstat} The Phelpsian perspective regards discrimination as purely a problem of statistical inference, where identical populations receive unequal treatment due solely to their access to different technologies to communicate their talents to potential employers.  \cite{arrow1973} points out that people's ability to convey their productivity to employers influences their choice to invest in their productivity in the first place. The core message of the literature following \cite{arrow1973} is that unequal treatment across identities can arise from ``self-fulfilling prophecies,'' whereby employers conjecture that employees' identities meaningfully signal some information about their productivity,  and, understanding this preconception, employees optimally  behave in a way that confirms the initial discriminatory conjecture. 

For example, in the model in \cite{coateloury1993}, firms wish to hire workers that are skilled; skill is acquired by a worker via a costly investment. At the moment of hiring (once workers have made their costly investments), to form a belief about the probability that a worker is skilled, the firm combines an observed signal with their prior conjecture of the overall proportion of workers in a population that invest in skill acquisition. For example, the firm may think that identity-$A$ workers invest in skill-acquisition in a  greater proportion than identity-$B$ workers. This is where the self-fulfilling nature of discrimination comes in: Because the prior conjectured by the firm affects their belief about a worker's skill, the  incentives to invest in acquiring skill are themselves affected by the firm's conjecture. \cite{coateloury1993} show that, because of the nonlinear relationship between the firm's prior and the workers' actions, there may be multiple equilibria, some in which agents in different populations (who are ex-ante equally able to acquire skill) choose different investment levels and are hired at different rates by the firm.

The Arrovian perspective highlights that investment incentives depend on how workers expect their investment in skill will be signaled to potential employers. Recently, \cite{onuchicray2023} propose a model where workers can signal their quality by working in teams. Individual's willingness to participate in certain teams depends on how they expect the credit for joint work will be perceived by the public (say, future potential employers). 

In the model, there are two potential ex-ante equal partners ($A$ and $B$), who each have an idea for a project and can choose to collaborate. These ideas can be better or worse, and their potential is correlated with the underlying quality of the worker that produced it. If the workers choose to not collaborate, they instead each work alone in their own idea; the public sees their separate project outcomes and from them infers the quality of each worker. If instead partners combine their ideas, then the public sees their joint project outcome, but not the idea contributed by each individual, and infers workers' qualities based on a conjecture of how much each individual provided to the collaboration. The equilibrium condition requires that the observer's conjectures about individual contributions are correct on average: credit attribution and the partners' collaboration strategies are co-determined in equilibrium. This co-determination is the mechanism behind the existence of multiple (perhaps discriminatory) equilibria.

To understand, suppose that the observer conjectures that partner $A$ contributes better ideas to joint projects than partner $B$ does. So, upon seeing a certain collaborative outcome, the public understands that more credit for that outcome should be assigned to person $A$ than to person $B$. Given a collaborative outcome, despite the partners being ex-ante equal, the observer forms a better posterior belief about $A$'s quality than of $B$'s quality. Anticipating this credit assignment, partner $A$ is indeed more willing to share good ideas with partner $B$, while partner $B$ is more likely to wish to keep good ideas to themselves, working alone so that credit for their good ideas are not scooped by partner $A$. 
This strategic response, in turn, can justify the observer's initial conjecture: indeed, person $A$ normally shares better ideas with person $B$ than vice versa, so that the correct inference is to assign more credit to $A$ than to $B$ when they collaborate. This discriminatory credit assignment is an equilibrium outcome, rather than an exogenous bias on the part of the observer -- it is \emph{self-fulfilling}, as in \cite{arrow1973}.\footnote{
\cite{onuchicray2023}  characterize productive environments where stable equilibria feature self-fulfilling discrimination -- for example, careers where reputational outcomes are very important, relative to the direct value of workers' projects, are relatively more susceptible to the discriminatory mechanism. The paper relates this observation to the empirical finding --- in \cite{sarsons2021} --- that there exists gender discrimination in credit assignment in the context of economics academia (a career where reputation-building is highly rewarded).}

\subsection{An Economic Perspective on Algorithmic Fairness} \label{sec:design4} Algorithms, rather than humans, are increasingly responsible for making decisions, or aiding decision-making, in many economically relevant contexts, such as household credit markets, bail judgements, and hiring. Research in computer science extensively documents settings in which algorithms fare better than human decision-making; but also raises alarm for the possibility that algorithmic decision-making can be ``unfair'' or discriminatory.\footnote{For a mostly empirical perspective on the mechanisms behind algorithmic bias, see \cite{cowgilltucker2020}. See also \cite{kasyabebe2021} for a discussion on what is considered (or should be considered) to be a biased or ``unfair'' algorithm.} 
Like humans, algorithms can be ``unfair'' for a variety of reasons: they may be designed with biased objectives, they may be exposed to data that is not equally representative of different populations, and may also provide ``self-fulfilling'' differential investment incentives to agents with different identities. As a consequence, many of the ideas introduced in the early models discussed above have been recently developed and adapted to an algorithmic decision-making context. Much of these recent developments take a very policy-oriented perspective, asking how algorithms should be developed and regulated when fairness is a policy goal.

\cite{lianglumuokumura2024} propose a model in which an algorithm performs a classification task: it observes a set of covariates about an individual (including maybe their population identity), which it uses to predict that individual's underlying type, and assign to it a ``best decision.'' The decision problem may be, for example, to grant bail or not to grant bail to the individual; in which case the relevant underlying type is whether the individual will show up to court if they receive bail. In this environment, the algorithm designer can have multiple goals, which perhaps conflict with each other. On the one hand, the designer wishes to put forth an algorithm that maximizes decision accuracy, minimizing errors that arise from misclassification, given the available data.\footnote{The model in \cite{lianglumuokumura2024}  allows for a wide class of ``error loss functions'' that describe the value of the assignment of types to decisions. These different loss functions map into different notions of accuracy and fairness as the designer's objectives.} On the other hand, they wish to maximize fairness, as measured by the difference in ex-ante expected loss yielded to individuals with belonging to one of the two possible identity groups. 

The paper provides a characterization of the fairness-accuracy frontier, delineating the combinations of fairness and accuracy values that are attainable by some algorithm  given the statistical environment generating the data.\footnote{\cite{dovalsmolin2024} propose a similar exercise, characterizing the set of welfare profiles attainable to an designer who designs information provision about underlying agents' types to a decision maker facing some decision problem --- perhaps a classification problem, as in \cite{lianglumuokumura2024}. Here a welfare profile describes the expected value to agents in different groups, so characterizing the set of attainable welfare profiles also encompasses a characterization of the fairness-accuracy trade-off. A more tangentially related problem is considered by a developing literature in mechanism design that studies the role of markets (and the design of markets) in welfare redistribution --- see, for example,  \cite{dworczakkominersakbarpour2021} and \cite{akbarpourdworczakkominers2024}.} By characterizing this frontier in terms of primitives of the data environment available to the algorithm designer, the authors can pose two types of novel questions: First, under what conditions on the data environment can we say that accuracy and fairness are objectives that are at odds with each other? Second, from the perspective of a regulator who can design the statistical environment --- perhaps by coarsening or altogether banning some inputs available to the algorithm --- what are gainful or harmful interventions?

Answers to both of these questions rely on a novel proposed property of the statistical environment: \emph{group-balance}, which holds if  the algorithm that yields the lowest error value to a group (that group's ``best algorithm'') yields a lower error to that group than to the other group. If the environment is group-balanced, maximizing fairness is an objective that is not at odds with the maximization of accuracy. If instead it is group-skewed,  there is a proper trade-off between these two goals. With respect to the second question. the paper shows that if the statistical environment is group-balanced and group identity is one of the available inputs, then excluding group identity or any other input that is informative about individuals' underlying types from the dataset strictly worsens the entire accuracy-fairness frontier --- which implies a decrease in the value to any algorithm designer, regardless of their weighting on accuracy and fairness objectives. 

\cite{strackyang2024} take a different perspective on fairness when analyzing a problem in which an information designer shapes the information about some underlying population that is made available to a decision maker who makes choices relevant to these individuals. Rather than considering fairness as a notion of closeness in value provided to members of different populations, they approach fairness as a requirement that decision-maker's choices based on the provided information be statistically independent of individuals' protected characteristics, such as population identity. This requirement imposes that the information designer must disclose information to the decision maker in a manner that maintains individuals' protected characteristics \emph{private}.\footnote{The idea of privacy preservation --- in the sense that decisions should be made statistically independent from protected characteristics --- is a common fairness criterion used in the computer science literature on algorithmic fairness.} 

\cite{strackyang2024} characterize the set of privacy-preserving signals that can be designed by the information provider. They show that privacy-preserving signals are built as the garbling of ``quantile signals.'' For example, suppose the information designer wants to convey information about individuals' credit scores, but is constrained to maintaining the privacy of individuals' gender identities. One possibility for that designer is to provide a signal that states that an individual ``belongs to quantile x of the credit score distribution among those with their same gender identity,'' without specifying to which gender identity that individual belongs. Because this signal is based on information about quantiles within the protected gender groups, and quantiles are by definition distributed equally across groups, it does not convey any information about individuals' gender identities. The paper also provides a characterization of optimal privacy-preserving signals for various classes of decision problems. In particular, the authors propose a two-step procedure that attains optimal privacy-preserving decision-making in a context that encompasses the classification problem considered in \cite{lianglumuokumura2024}\footnote{First, an algorithm should be asked to most efficiently generate a score for each individual based on all available information (where score here is the predictive variable relevant to the decision problem at hand), without concern for fairness or privacy; next, the generated scores go through a ``post-processing step'' which computes the quantile signals of the predicted scores for each group protected characteristic. These quantile signal scores are the only information to be provided to the decision maker, thus ensuring that privacy is preserved.}

In a context more closely related to Arrow's approach to statistical discrimination, \cite{zhu2024} considers the regulation of algorithms towards attaining non-discrimination/fairness. Remember that in \cite{coateloury1993}, agents in different groups may have different incentives to invest in acquiring a skill due to their expectation that future employers learn about their skills differently. \cite{zhu2024} approach posits that a set of policies ensures fairness if they imply that any equilibrium of the game subject to those policies is one in which all groups have the same incentives to acquire skill. 
The policy interventions considered by \cite{zhu2024} are especially targeted at regulating algorithmic decision-making, as opposed to decision-making done by humans. These two contexts differ in an important way: the ``computed beliefs'' calculated by a machine towards recommending a decision are verifiable, while beliefs computed by the human brain cannot be audited. \cite{zhu2024} argues that policy interventions that target these verifiable beliefs are able to attain the proposed notion of fairness; in contrast, if beliefs are not verifiable and cannot be targeted by regulatory policies, fairness, in the sense of equalizing investment incentives, might be an unreachable goal. 

The paper proposes a policy intervention that restricts decision-makers to apply the same acceptance standard across groups, implemented through what \cite{zhu2024} calls \emph{common identity}. Concretely, the intervention requires that acceptance or rejection decisions be based on a single score constructed by aggregating the algorithm's identity-specific predictions, and that a common cutoff be applied to this score for all groups. In this sense, the policy is similar in spirit to privacy-preserving quantile signals, as in \cite{strackyang2024}: while identity may enter the construction of the score, final decisions are constrained to be independent of group membership. \cite{zhu2024} shows that, under this intervention, discriminatory equilibria are eliminated even in settings where standard outcome-based policies fail, and that the result is robust to biased beliefs on the part of the decision-maker.

\section{Learning and Discrimination}
\label{sec:2}
Theories of statistical discrimination (in the context of labor markets) rely on the basic assumption that, at the time of hiring, employers imperfectly observe the aptitude of different candidates for the job position being offered. Recent contributions propose models that give more structure to the employer's problem of assessing employees' productivity, often posing the acquisition of information about employees as an endogenous decision. An important new insight is that following an optimal information acquisition policy can often lead employers into \emph{learning traps}, in which they acquire more or better information about employees with ``favored identities.''

\subsection{Learning Traps and Discrimination}
\label{sec:screening}
\cite{bardhiguostrulovici2024} propose a model where time is continuous and, at each instant, a firm assigns to a task at most one of two workers of unknown skill. Each worker $i\in\{A,B\}$ is either of high or low quality, with $p_i\in(0,1)$ being the initial prior probability that worker $i$ is of high quality. Worker $A$ is assumed to have ex-ante higher expected quality, so that $p_A>p_B$, but the paper is mainly interested in the case where workers are almost equal, so that $p_B\uparrow p_A$. The firm wishes to employ either worker if and only if they are of high quality. 

To learn about a worker's quality, the firm must employ them and watch their performance at the task: If a worker is employed at the time interval $[t,t+dt]$ and their type is $\theta\in\{h,l\}$, then a public signal arrives with probability $\lambda_\theta dt$. This specification encompasses two cases of interest: breakthrough learning, where $\lambda_h>0=\lambda_l$, so that an observed signal reveals that the worker is of the high type; and  breakdown learning, where $\lambda_l>0=\lambda_h$, so a signal reveals a worker's low type.
These different learning environments are meant to represent tasks performed in different types of firms. For example, scientific research can be thought of as a breakthrough learning environment, where most observed news -- say, a new published paper or an awarded grant -- are positive news about the researcher's underlying type. On the other hand, a nightwatch or an airline pilot work in breakdown environments, where publicly observed news are usually negative -- for example, a successful robbery attempt or an emergency landing. \cite{bardhiguostrulovici2024} show that, depending on the underlying learning environment, a small difference in ex-ante expected quality between worker $A$ and worker $B$ can lead to large differences in career trajectories and worker payoffs.

Consider the breakthrough learning environment. Because $p_A>p_B$, at the beginning, the firm allocates worker $A$ to the task. If a breakthrough is observed, then the firm learns that worker $A$ has high quality and allocates them to the task forever. Otherwise, at each instant where no breakthrough is observed, the firm's posterior that worker $A$ is of high type is updated downwards. This happens until some time $t^*$, where this posterior equals to the prior on worker $B$. From that point onwards, $A$ and $B$ are treated symmetrically, so that their expected career paths and payoffs coincide. If workers are almost equal ($p_B\uparrow p_A$), $t^*$ is very small, so that the probability of a breakthrough for worker $A$ in the interval $[0,t^*)$ is negligible. Thus from an ex-ante perspective, worker $A$ and $B$ have almost equal expected career paths. In this sense, the breakthrough learning environment is such that early-career discrimination --- due to the firm starting by employing worker $A$ --- does not perpetuate; it is self-correcting.

Equally in the breakdown learning scenario, the firm starts by allocating worker $A$ to the task; in case of a breakdown, $A$ is revealed to be of low type and is never again employed by the firm, who instead switches to employing worker $B$. If a breakdown does not happen, the firm's posterior on $A$'s quality is only ever updated positively, so that worker $B$ is never employed and no learning about their type ever takes place. In this case, early-career discrimination is spiralling: even when workers are almost equal ($p_B\uparrow p_A$), $A$ and $B$ have very different expected career paths, and the ratio between the expected payoffs of workers $A$ and $B$ does not approach $1$. This result shows that that discrimination can be a path-dependent and cumulative process: despite both employees being symmetric, the employer fails to learn or significantly delays learning about the worker with a dis-favored identity while they are ``trapped'' learning about the favored worker.

\cite{chekimzhong2019} show how discriminatory \emph{learning traps} can arise in markets where ratings and recommendations facilitate social learning among users. They introduce a model in which buyers and sellers repeatedly and bilaterally meet to trade. Sellers differ in terms of the quality of the good they provide, good or bad, and also have differing payoff-irrelevant identities. There is a platform that provides buyers with ratings that are informative about the quality of the goods provided by each of the sellers. The platform is assumed to be unbiased in the sense that, given some information about a seller's quality, it generates the same rating regardless of that seller's identity. 

Discrimination may still arise due to the platform's data-acquisition process: information about a seller's quality is generated whenever they transact with a buyer, who then reports to the platform whether the good they received was of low or high quality. Because data is sampled only when transactions occur, and because buyers wish to transact with the sellers they believe are of high quality -- those with good ratings -- then much data is generated about sellers who already have high ratings, and little data is observed about low-rated sellers. 
A discriminatory feedback loop may ensue: if sellers of a dis-favored identity are seldom sampled by buyers, despite their good ratings, then a good rating becomes a less informative signal about their quality. Which, in turn, encourages buyers in the next period not to transact with positively-rated sellers of the dis-favored identity. 

In \cite{chekimzhong2019}, discrimination arises as a learning failure due to the collective of buyers acquiring information about sellers through the platform. The paper then suggests that such learning failures can be avoided through careful design of the algorithm behind the ratings shared in the platform, so as to incentivize more equitable sampling of all buyers. Similar policies are suggested by 
\cite{komiyamanoda2024}, who consider an environment where discrimination similarly arises as a failure of social learning. In their model, a sequence of myopic firms decide to hire a worker from a set of available candidates (the multiple ``arms'' in a bandit problem). Each firm wishes to hire the most skilled worker, and they infer workers' skills from some observable worker characteristics, which include the worker's identity $A$ or $B$. At first, no firm knows precisely how to interpret these characteristics, but they learn the statistical association between characteristics and skills using data pertaining to past hiring cases.
  
The paper introduces the possibility of a phenomenon called \emph{perpetual underestimation}: Suppose at some point in the stochastic learning process, workers with a identity $A$ have their skill underestimated. They then appear to be a risky or undesirable hiring choice for a firm, who instead might prefer to hire a worker with identity $B$. At this point, a learning trap ensues: as long as firms only hire $B$ workers, society cannot learn about the workers in group $A$, and the imbalance persists in the long run. Perpetual underestimation arises due to each firm's myopic behavior. In each period, the myopic firm chooses the safe arm and hires a worker with the best record given the current information. By doing so, the firm forgoes exploring the risky arm, which would generate more information about the underestimated population, to be used by future firms. This perpetual underestimation phenomenon is not only ``unfair,'' but can be inefficient. \cite{komiyamanoda2024} show that either using a exploration-subsidy mechanism or temporarily using the Rooney rule, which requires each firm to interview at least one minority candidate as a finalist for each job opening, can effectively mitigate discrimination caused by insufficient data, as well as improve welfare.

While each of these papers proposes a theory based on learning traps, \citeauthor{liraymondbergman2024}'s \citeyearpar{liraymondbergman2024} work provides a useful empirical illustration of the exploration versus exploitation trade-off in the multi-armed bandit problem faced by hiring algorithms: to find the best workers over time, firms can exploit the safe bandit arm, by selecting from groups with proven track records, or explore the risky arm by selecting from under-represented groups to learn about their quality. Standard, exploitative hiring rules --- akin to supervised learning algorithms --- can perpetuate discriminatory learning traps by repeatedly selecting from groups about which a firm already has precise information. By contrast, hiring algorithms that explicitly incorporate exploration give greater weight to candidates whose predicted performance is more uncertain, which tends to increase sampling of historically under-represented groups. Using data from hiring process in a Fortune-500 firm, the authors show that such exploration-oriented algorithms can mitigate discrimination while also improving hiring outcomes over time, highlighting how discrimination can arise from myopic learning rather than biased preferences.\footnote{In the same spirit, \cite{lepage2024} and Benson and \cite{bensonlepage} test ``learning trap discrimination'' predictions using a labor market experiment (the former paper) and administrative records from a large national US retailer (the latter paper). \citeauthor{lepage2024}'s \citeyearpar{lepage2024} results illustrate the formation of biased beliefs based on employer's experience with previous workers. And Benson and \citeauthor{bensonlepage}'s \citeyearpar{bensonlepage} results suggest that managers develop biased beliefs from endogenous learning about racial groups, thereby systematically disadvantaging minority workers.}



\subsection{Learning with Costly Information Acquisition}
\label{sec:inat}
The models introduced in section \ref{sec:screening} assume that information about a worker or a seller is only produced when they are employed or a transaction occurs. This section presents a different perspective, according to which costly information is actively and flexibly sought out by decision makers.

\cite{bartosbauerchytilovamatejka2016} study a model where a \emph{rationally inattentive} employer can learn a potential employee's quality to any degree of precision, but at some increasing attention cost. For example, they can gather a lot of information before making a decision, by calling the candidate's previous employers, reading their resum\'{e} carefully, and having many people interview the applicant. Alternatively, they could simply skim the provided resum\'{e} before making an, admittedly less informed, decision.
The candidate belongs to either identity $A$ or identity $B$, and identity $A$ is assumed to contain candidates that are on average more qualified than those of identity $B$. The candidate's identity is observed by the employer before they decide how much attention to put towards screening -- think of an employer seeing the candidates' names right at the beginning of a selection process for a job. The main result in this paper argues that the employer's rational inattention often amplifies the expected outcome differences between individuals of the two identities.\footnote{\cite{bartosbauerchytilovamatejka2016} also conduct field experiments and find that, in various contexts, the attention decisions of potential employers and landlords are consistent with their model's predictions.} 

This amplification arises through different mechanisms depending on the selectivity of the hiring process. A hiring process is highly selective if, without observing any extra information, the employer would choose not to accept an application from either a candidate of identity $A$ or a candidate of identity $B$. The authors call this a cherry-picking market. In this case, the rationally inattentive employer optimally pays more attention to an application from a candidate of identity $A$ than one from a candidate of identity $B$. Consequently, candidates of identity $A$ who are of high-enough quality are often recognized as such, and accepted by the employer. Conversely, a candidate of identity $B$ with the same, high-enough, quality is less likely to be properly identified, and thus more often assigned the default rejection. 
In other words, the optimal attention assignment implies that the probability of type-2 errors, wherein a good candidate is not hired, is lower for identity $A$ than for identity $B$. This mechanism amplifies the difference between  the ex-ante hiring probability of candidates of identities $A$ and $B$, relative to a benchmark where the same attention level is paid to applications coming from both identities.

In contrast is a lemon-dropping market, where absent any extra information, the employer's default action is to accept a candidate from both identities. The rationally inattentive employer now optimally pays more attention to applications from identity $B$ than to those coming from members of identity $A$. As a consequence, candidates of identity $B$ who are not of high-enough quality, are more likely than their identity-$A$ counterparts to be recognized and accordingly rejected. In this case, the probability of type-1 errors, wherein a bad candidate is hired, is lower for identity $B$ than for identity $A$. Once again, this mechanism is discriminatory, in that it amplifies existing differences between the two identities. 

\cite{mcgee20252} considers a rationally inattentive learner, who is initially exposed to an impulse about the quality of a possible hire, determined by their own past experiences or their surrounding social setting. The learner chooses between following this initial impulse or incurring in costly deliberation to learn the true quality of the potential hire. The author explores a dynamic environment, where past experiences influence future impulses, and shows that long-term discrimination is more prevalent in settings where impulses are formed from second-hand influences, instead of solely one's own experiences. 

\cite{fosgerausethiweibull2023} study a static model with a  rationally inattentive screener, observing that the attention decision of the screener itself determines the incentives for workers of different identities to invest in acquiring skill in the first place. The codetermination of skill-choice and screening-attention imply that there are multiple equilibria -- similarly to \cite{coateloury1993} -- and ex ante identical categories can receive asymmetric equilibrium treatment.
\cite{echeniqueli2024} also propose a model where a principal rationally acquires information about the skills of agents in majority and minority groups. By rationally allocating scarce attention between the groups, the principal may create incentives for majority agents to invest in skill acquisition, and for minority agents not to do so. Interestingly, \cite{echeniqueli2024} show that as the attention cost decreases, the difference in skill acquisition (and payoff) between the majority and minority groups may increase; thereby highlighting that ``statistical'' discrimination may indeed worsen as information becomes more readily available.


\section{Discrimination with Mispecified Models and Other Learning Heuristics}
\label{sec:3}

The work discussed in this section departs from classic model of statistical discrimination in that they consider learning processes other than Bayesian updating. Specifically, these papers consider decision-makers that learn through Bayesian updating with mis-specified models, and other heuristics such as the representative heuristic, and learning with a ``self-image bias.'' These papers propose methods to identify discrimination due to such learning heuristics and propose policies that address discrimination stemming specifically from these behavioral biases.

\subsection{Identifying Mis-Specified Discriminatory Learning}
\label{sec:inac}

\cite{bohrenimasrosenberg2019} propose a model where a sequence of evaluators tries to learn about an agent's underlying ability based on their reported performance on a series of tasks (reported by previous evaluators), as well as on their group identity. 
An agent belongs to a group $g\in\{A,B\}$ and has unobserved ability $a$, drawn from a normal distribution, with mean  potentially distinct across groups, and some group-independent precision. The agent sequentially performs two tasks (in periods 1 and 2, for a simplified version of their model), each of which generates a performance outcome equal to $a$ plus some normally distributed, mean-zero, error. 

In each period, a short-lived evaluator makes a report, assessing the agent's ability. In period 1, the first evaluator observes only the agent's performance outcome generated by that period's task before making their report. In period 2, the second evaluator sees period 2's task performance, as well as the first evaluation. In making their reports, evaluators wish to precisely estimate the agent's underlying ability. 

Each of the evaluators is ``mostly Bayesian'' when estimating the agent's underlying ability, with two caveats. First, they might choose to report an evaluation that is worse than the true ability estimate when the agent belongs to a group the evaluator is biased against. Second, the evaluator may form this evaluation through a mispecified model of the world. There are two potential misspecifications: (i) the evaluator may hold a wrong belief about the average ability among members of a group; and (ii) an evaluator may be wrong about the distribution of beliefs and preferences of other evaluators. The second potential mispecification is relevant to the period 2 evaluator who must interpret the report made by the evaluator in period 1 before making their own assessment of the agent's ability. 

Group $g\in\{A,B\}$ is said to be discriminated against in period 1 after a certain performance outcome if the average report made by period-1 evaluators after seeing the outcome for an agent in group $g$ is smaller than the average report made after seeing that same outcome coming from an agent of the other group. 
Group $g\in\{A,B\}$ is said to be discriminated against in period 2 after a given period-1 evaluation and a given period-2 performance outcome if the average report made by period-2 evaluators after seeing this history for an agent in group $g$ is smaller than the average report made after seeing that same history coming from an agent of the other group. 

Two theoretical results in the paper propose the presence or absence of period-1 and period-2 discrimination as identifying the source of discrimination --- bias, correctly-specified belief-driven discrimination, or belief-driven in a mis-specified learning environment. The first result shows that, if discrimination is belief-driven, then period-1 discrimination is decreasing in the accuracy of the period-1 performance outcome. The same is not true if discrimination is due to bias.
The second result concerns the dynamics of discrimination. Suppose there is period-1 discrimination against group $A$, and suppose it is belief-driven. If all period-2 evaluators have correctly specified models, then there cannot be any period-2 discrimination against group $B$ -- that is, there is \emph{no discrimination reversal}. If instead some  evaluators hold wrong beliefs and underestimate the average ability in group $A$, then there may be period-2 discrimination against group $B$. That is, the presence of some mis-specified evaluators can generate discrimination reversal between periods 1 and 2.\footnote{\cite{fryer2007} also points out that discrimination reversal may take place in a dynamic model of statistical discrimination, in a model in which the belief updating process is assumed to be correctly specified. The main observation in \cite{fryer2007} is that, if at the moment of hiring, an employer discriminates against members of group $A$, then, conditional on being hired, a member of group $A$ is relatively more talented than a member of group $B$. Consequently, if the employer then wishes to pick a member of the hired group to promote, they may favor members of group $A$ who were discriminated against in the initial stage. 

The main extra ingredient in the model of \cite{fryer2007}, relative to \cite{bohrenimasrosenberg2019}, is that at each stage workers can invest in their productivity -- think of \cite{fryer2007} as a repeated version of \cite{coateloury1993} model. In the first stage, discrimination against group $A$ is a self-fulfilling equilibrium, where group $A$ members invest less in acquiring skills. But, conditional on a person $A$ being hired, the employer expects that they are more likely to have a low cost of skill acquisition, and thus expects that they will acquire further skill in the second stage, before a promotion decision. As such, a group-$A$ worker indeed has greater incentives to make a second-stage skill investment, and is thus relatively favored by the employer.
}

To understand this result, first note that the period-1 discrimination against group $A$ is straightforwardly driven by the presence of some evaluators who underestimate group $A$'s average ability. The period-2 reversal happens because period-2 impartial evaluators, after seeing a period-1 report about a group-$A$ member, understand that it may have been made by a mis-specified evaluator, who would have under-reported. As such, a period-2 impartial evaluator interprets a period-1 report about a group-$A$ member more favorably than if it referred to a group-$B$ member. Consequently, impartial period-2 evaluator discriminates against group $B$. If there are enough impartial period-2 evaluators, who themselves believe there are enough other evaluators that underestimate group $A$, there is period-2 discrimination against group $B$ on average.

Having introduced these results that can identify sources of discrimination, \cite{bohrenimasrosenberg2019} then put them to use in an experiment run in an online platform where users post questions and answers and also rate each other's questions and answers.\footnote{\cite{bohrenhaggagimaspope2023} propose an alternative experiment and  methodology to identify the presence of discrimination stemming from mis-specified beliefs and learning; in it, beliefs of ``employers'' are directly elicited in a laboratory setting.} In the experiment, they generate new accounts on this platform, varying the gender ascribed to them. First, they explore how the rating of female versus male accounts differ for recently generated accounts. They show that men are significantly more rewarded (in terms of ratings) for posting good questions than women are; and that the reward for good answers does not significantly differ across genders. The authors argue that posted answers are a more precise measure of a user's ability than their posted questions; and so they conclude that ``period-1 discrimination'' decreases with the precision of the test. This is consistent with discrimination being belief-based, rather than due to bias. Next, the authors compare ratings for novice accounts versus ratings for accounts that have an established track record. This comparison shows evidence of ``discrimination reversal,'' which is consistent with discrimination driven by mis-specified beliefs, as argued in their theory.

\cite{bohrenhullimas2023} make a further remark on the phenomenon of discrimination reversal, which leads to them proposing a further decomposition of discrimination into a \emph{direct} and a \emph{systemic} component. This decomposition refers to a large body of work in other social sciences that views discrimination as a systemic phenomenon, viewing disparate group-based treatment as a ``cumulative outcome of both direct and indirect interactions between outcomes and evaluations across different stages and domains.'' Suppose, as in the discrimination reversal example, that there is a  hiring process consisting of two stages. In a first stage, a recruiter meets a potential hire and issues an evaluation. This recruiter is biased and consistently evaluates female candidates less favorably than equally able male candidates. At a second stage, a hiring manager at the firm observes the recruiter's evaluation and makes a hiring decision which, conditional on the first stage evaluation, favors female candidates. An econometrician who is only able to observe the second stage of the hiring process may conclude from the data that there is discrimination in favor of female candidates. \cite{bohrenhullimas2023} instead suggest that the econometrician should conclude that ``there is direct discrimination by the hiring manager,'' but the statistical finding does not rule out \emph{systemic discrimination} against female candidates in the hiring process as a whole. Indeed, we know that the first recruiter consistently under-evaluates women and so, if the econometrician observes the hiring process as a whole, they may find that women are less likely to be hired, when compared with equally able men. In the paper, \cite{bohrenhullimas2023} more broadly define the distinction between systemic and direct components of discrimination and propose an econometric methodology to measure this decomposition.

\subsection{Misperceptions in a General Statistical Discrimination Setting} The statistical discrimination setting introduced in \cite{bohrenimasrosenberg2019} is a slight variation from Phelps' original model, introducing the possibility of mis-specification, but maintaining that agents' ability types and the errors in the signals about their abilities are normally distributed. \cite{escudeetal2025} consider a general model of statistical discrimination, in the spirit of \cite{chambersechenique2021}, in which firms may \emph{mis-perceive} the true skill distribution in a population. Specifically, the authors posit that the \emph{prior} type distribution in a population may differ from observing firms' \emph{perception} about that type distribution. The main goal of the paper is to investigate the relationship between more informative signals and better average pay, going back to \citeauthor{aignercain1977}'s \citeyearpar{aignercain1977}, but in a context with prior misperceptions. 

The paper’s central contribution is to disentangle two distinct channels through which additional information (a more informative signal) affects average pay in a population. The first is an instrumental channel: more informative signals allow firms to more profitably employ workers, and therefore improve their wages. The second is a perception-correcting effect, through which new information reduces the influence of firms' perceptions about the skill distribution in the population. When a population is initially under-perceived, additional information raises average pay by correcting pessimistic beliefs; when it is over-perceived, the same informational improvement lowers average pay by eroding overly favorable perceptions. 

The authors apply this decomposition to clarify the circumstances under which better information decreases the gap in pay between two populations that are misperceived --- remember that in \citeauthor{bohrenimasrosenberg2019}'s \citeyearpar{bohrenimasrosenberg2019} Gaussian setting, better information always decreases pay gaps.\footnote{More generally, much of the experimental literature on discrimination uses ``information interventions'' that improve the informativeness of signals with the goal to reduce discriminatory gaps. See \cite{litwinlow2025} (section 2.4) for a general reference to these information interventions, and \cite{laoenanrathelot2022} and \cite{cuilizhang2019} for specific implementations of that intervention, in the context of ratial discrimination on Airbnb.} More generally, \cite{escudeetal2025} find that pay gaps driven by misperceptions decreases following an improvement in information, provided a set of stringent requirements is satisfied. In turn, better information can often backfire and actually increase  gaps.

\subsection{Stereotypes}
\label{sec:stereo} The work described in the previous two sections departs from classic theories of discrimination in that it considers a more general process through which decision-makers learn about agents of different identities: mostly Bayesian updating, but perhaps with mis-specified prior beliefs about the ability distribution in different populations (and about the learning process of other decision-makers). An alternative departure from the Bayesian learning setting posits that the belief updating process is affected by the learner's ``stereotyped'' understanding of members of different groups. 


\cite{bordalocoffmangennaiolishleifer2016} propose a model of stereotypes based on the representativeness heuristic due to \cite{kahnemantversky1972} and \cite{tverskykahneman1983}. This heuristic describes an individual belief-formation process that may lead to inaccurate beliefs. This representativeness heuristic anchors itself on salient identity traits, such as gender or racial identity, and may thus be useful for understanding belief differences attributed to these different groups.

Take two groups $g\in\{A,B\}$, and evaluate the distribution of certain traits $t\in T$ in each group: let $\pi_g(t)$ be the probability that trait $t$ is present in group $g$. It is useful to think of the traits in set $T$ as mutually exclusive, such that each member of group $g$ has exactly one of the traits -- for example, $t$ could be a number of schooling years, and $T$ the set of possible schooling years. The representativeness of trait $t$ in group $g$ is
\begin{equation}
\label{eq:3}
R(t,g)=\frac{\pi_g(t)}{\pi_{-g}(t)},
\end{equation}
where $-g$ is the group that is not $g$. They posit that, while people understand the distributions $\pi_A$ and $\pi_B$, these are not the distributions that are salient to them when making assessments about groups $A$ and $B$. Rather, they use  ``stereotyped'' distributions $\pi^{st}_A$ and $\pi^{st}_B$, which overweight the traits that are representative of groups $A$ and $B$, respectively -- as measured by their representativeness given by equation (\ref{eq:3}).

The main results in \cite{bordalocoffmangennaiolishleifer2016} assess how small differences in trait distributions across groups can be exacerbated by the stereotyping heuristic.\footnote{They also provide experimental and observational evidence that this heuristic is a good approximation for people's belief formation processes. \cite{bordalocoffmangennaiolishleifer2016} specifically assesses the stereotyped belief formation process across genders -- both by members of different gender groups, and across members of different gender groups.} It is important to note that, while the behavioral economics literature has recently proposed models of a variety of belief-formation heuristics, this stereotyping behavior is particularly relevant to questions of discrimination, because it is based on the perceived distinction between groups $A$ and $B$ in the first place. 

Recently, \cite{espondaopreayuksel2023} propose a model where decision-makers employ the same representativeness heuristics not to the prior type distributions in populations $A$ and $B$, but rather to new information they learn about agents in these populations. They introduce a novel cognitive bias denoted ``representative signal distortion,'' according to which agents misinterpret new information about an agent to be more representative of the agent's group, in contrast to a reference group. The authors show in an experimental setting the effectiveness of two policies at correcting this cognitive bias: presenting information to subjects before they learn about agents' group-membership; and asking subjects to make decisions about agents of a single group, so that subjects are prevented from contrasting different groups.

\subsection{Self-Image Bias} \cite{heidhueskoszegistrack2019} also propose a behavioral model of interpretation of individual outcomes and show that identity-based interpretation of individual outcomes can result from an individual's mis-perception (overconfidence) about their own ability. 
In their model, society is comprised of $K$ potentially overlapping groups, and each agent is either a member, a competitor, or a neutral outsider of a group. An agent observes the ``recognition outcomes''  of all members of society, including their own (for example, people's achievements or social statuses). These outcomes are a result of each person's underlying ability type, but also some noise. Moreover, the agent posits that there may be some discrimination in place, benefiting members of a group, and hurting competitor groups.

The agent's objective is to learn the true discrimination pattern in society based on his observations (the agent is a Bayesian learner). The one crucial behavioral assumption in the model is that the agent is ``stubbornly overconfident:'' they hold a \emph{point belief} about their own ability, which is \emph{above the correct one}. The authors show that adding this one behavioral element to an otherwise standard model generates a series of empirically verified patterns in social beliefs.

The first implication of the agent's biased learning process is that, in the long-run (after gathering a lot of data on recognition outcomes), they overestimate societal discrimination against any group they are a member of, and underestimate it against any group they are in competition with. This conclusion is reached as an explanation for their own recognition outcomes, which fall short of their overconfident expectations -- the stubbornness of the agent's overconfidence implies that they update their beliefs about discrimination patters, rather than review their beliefs about their own underlying ability. 
An almost immediate implication of this first result is that the agent holds favorable views about the ability of people that belong to the same groups as their own, relative to their recognition outcomes. A converse to this ``in-group bias'' is that the agent holds overly disfavorable views about the ability of people in competing groups.


Like \cite{heidhueskoszegistrack2019}, \cite{siniscalchiveronesi2021} propose a model where discrimination is a consequence of agents having a \emph{self-image bias}.\footnote{See also \cite{mcgee20251}, where individuals form self- versus other-regarding stereotypes from motivated reasoning when discriminating can produce material economic gains.}
The model in \cite{siniscalchiveronesi2021} is applied to the academic profession: there is an overlapping-generations environment where established researchers evaluate new researchers. Each researcher -- new or established -- belongs to one of two groups, say $A$ and $B$, and their group membership is not payoff relevant in itself. Beyond their group membership, researchers are also heterogeneous with respect to the set of characteristics they are endowed with -- for example, whether they are theoretical or empirical researchers, or whether they are interested in macro- or micro-economic questions. The distribution of characteristics differs across groups, but importantly all characteristics have the same positive effect on the likelihood of high-quality research, so that both groups are ``equally qualified.''

In each period, each new researcher produces research of some quality that depends stochastically on their own underlying characteristics. The research quality is then perfectly observed by one member of the established generation (the referee), who decides whether or not to accept the young researcher as a member of the established population. This is where the self-image bias comes in: each referee, in deciding whether to accept the young researcher, cares not only about their research quality, but also about whether the young researcher's characteristics match their own. Accepted researchers become established in the next period, and thus referees of future cohorts, and researchers who are rejected leave the model. 

In that environment, \cite{siniscalchiveronesi2021} study the dynamics of the population of academics. Their main observation is that, in the presence of self-image bias, even mild between-group heterogeneity generates a persistent bias in favor of young researchers who belong to the initially larger group (say, group $A$). Beyond that, even though there are successful group-$B$ researchers, they are more likely to be those whose characteristics are close to the ones more prevalent in group $A$. An interesting implication is that the pro-group-$A$ bias is in fact perpetrated even by established group $B$ researchers, because these successful $B$ researchers were endogenously selected to be ``closer'' to common group-$A$ characteristics. 

\subsection{Other Heuristics} \cite{hubertlittle2023} propose a theory of ``discrimination in policing,'' stemming from a learning heuristic they denote non-conditioning bias -- akin to correlation neglect, as in \cite{ortolevasnowberg2015} and \cite{levyrazin2015}. In \cite{hubertlittle2023}, a police department must decide how to allocate policing resources between different communities. To do so efficiently, they must form beliefs about the amount of crime to be expected in these different locations. When forming these beliefs, the officers misinterpret crime statistics: they do not properly account for the fact that they will detect more crime in more heavily policed communities. This creates a feedback loop, whereby communities that are over-policed generate crime statistics that are then misinterpreted, justifying the initial decision to over-police them. 

\cite{cunninghamdequidt2024} propose a model of bias based on a decision-making heuristic, rather than a learning heuristic. They introduce a distinction of biased decision making between that reflecting \emph{explicit} and \emph{implicit} bias in the decision-maker's preferences. In their leading example, they consider a hiring manager who always chooses to hire a woman over a man with the same qualifications, but always chooses to hire a man over a woman if their qualifications differ. They interpret such observation as evidencing an explicit bias of the manager for women --- whom they hire when the woman and the men are directly comparable --- but an implicit bias for men, who are hired when the two candidates' qualifications are not directly comparable. This behavior is interpreted as coming from a hiring manager who inherently wants to favor men, but is afraid of ``getting caught'' doing it; such a manager is more willing to act based on their ``true'' underlying bias when decisions are not immediately comparable by a conjectured audience.

\cite{sullivanmeyerbuchmann} introduce the idea that discrimination may be caused by other-regarding preferences. A firm, even if not subject to any behavioral learning constraints, may have perceptions about the payoff to individuals in two different groups which differ from their true payoff structures, and from each other. Such a firm may make \emph{paternalistic} decisions, treating individuals in the two groups differentially, even if against their will, to protect them (and especially members of one particular group) from what is perceived to be harmful or unpleasant situations.\footnote{Their paper focuses on labor-market gender discrimination: ``In the labor market, paternalistic discrimination may lead employers to hire women over men for female-stereotyped jobs, to avoid promoting recent mothers to reduce workloads, or to fire single workers over workers with families.'' }

\section{Conclusion}
This survey has reviewed recent theoretical contributions to the study of discrimination, which cut across the traditional taste-based versus statistical discrimination distinction. While that dichotomy remains useful as a historical reference point, many of the papers surveyed here are structured instead around features of learning environments, information structures, and belief formation processes. In this sense, recent theories are less concerned with classifying discrimination by motive, and more focused on understanding how disparate outcomes emerge from the structure of inference and decision-making problems.

A prominent theme in this literature is the growing role of algorithmic decision-making, which both motivates new theoretical questions and expands the set of feasible policy interventions. Several papers reviewed here analyze discrimination in environments where decisions or beliefs are computed by algorithms, allowing researchers to study how data, signals, and decision rules can be regulated directly. These contributions draw heavily on advances in information economics and information design. Many of the theoretical developments surveyed are also closely connected to broader recent work on learning with non-Bayesian or mis-specified agents, and are often motivated or complemented by laboratory and field experiments documenting biased learning, learning traps, and the effects of information and algorithmic interventions. While this article has focused on theory, this parallel empirical work has played an important role in shaping the questions and modeling choices in the recent theoretical literature

In the online appendix, I discuss a set of papers that I interpret as  discussing economic theories of institutional discrimination, taking the view of social norms as informal institutions and incentive mechanisms as formal institutions.

\clearpage

\appendix
\centerline{\Large\textbf{{\color{paradise}Online Appendix on Institutional Discrimination}}}
\vspace*{.5in}


In a paper published in the \emph{Journal of Economic Perspective} in 2020, sociologists Mario Small and Devah Pager criticize the economic research agenda on discrimination, arguing that it misses ``what sociologists and others have called `institutional discrimination,' `structural discrimination,' and `institutional racism,' which are all terms used to refer to the idea that something other than individuals may discriminate by race.'' 
In their essay, they define \emph{institutional discrimination} as ``differential treatment that may be caused by organizational rules or by people following the law,'' and say that it need not result from personal prejudice or from rational guesses on the basis of group characteristics.

To introduce the economic perspective on discriminatory institutions, we must first understand how economists view institutions. The Nobel-prize-winning economist Douglass North, in his \emph{Journal of Economic Perspectives} article in 1991, titled ``Institutions,'' writes (emphases and cuts are my own): ``Institutions are the humanly devised constraints that structure political, economic and social interaction. They consist of both \emph{informal constraints} (sanctions, taboos, customs, traditions, and codes of conduct), and \emph{formal
rules} (constitutions, laws, property rights). (...) Together with the standard constraints of economics they define the choice set and therefore determine transaction and production costs and hence the profitability and feasibility of engaging in economic activity. (...) Institutions provide the incentive structure of an economy; as that structure evolves, it shapes the direction of economic change towards growth, stagnation, or decline.'' 

The articles I survey in Sections \ref{sec:norms} and \ref{sec:design} exactly speak to the two types of institutions referred to by North -- models of \emph{informal constraints} (social norms) in the former and \emph{formal rules} (institutional design) in the latter. 
Despite my allusion to these theories as institutional discrimination, I do not believe them to fully (or even to a large extent) describe the sociological perspective on institutional discrimination. Rather, they are some notable examples of economic theories of discrimination that come closer to that approach. 

\section{Discriminatory Social Norms} 
\label{sec:norms}
An often overlooked economic ``theory of discrimination,'' which departs from the taste-based versus statistical dichotomy, is one that views disparate treatment of different identities as being a social norm. 
According to \cite{young2015}, ``Social norms are patterns of behavior that are self-enforcing within a group: Everyone conforms, everyone is expected to conform, and everyone wants to conform when they expect everyone else to conform. Social norms are often sustained by multiple mechanisms, including a desire to coordinate, fear of being sanctioned, signaling membership in a group, or simply following the lead of others.''

The first class of models I consider (section \ref{sec:norm1}) are roughly embedded in the environment of \cite{kandori1992}, and view discrimination as a social norm enforced in communities.\footnote{The model of community-enforced social norms in \cite{kandori1992} is not the only economic approach to modeling social norms. Later in this section, I discuss evolutionary game theory models of social norms following \cite{young1993}. Another approach is one that directly models agents' incentives to ``conform,'' as in \cite{bernheim1994}, or to conform to expectations put on their own social position, as in \cite{akerlof1997} and \cite{akerlofkranton2000}. Some work --- discussed for example in \cite{jackson2008} --- microfounds these theories of conformity by modelling the formation and evolution of social norms in networks.}  Section \ref{sec:stable} describes an evolutionary game theory approach to social norms, and comments on the stability of discriminatory norms. Both approaches conform with view in \cite{young2015} that a discriminatory social norm is sustained by a society's desire to coordinate, and by society members' fear of being sanctioned.
In section \ref{sec:norm2}, I briefly comment on a literature stemming from \cite{akerlofkranton2000}, which proposes that group-dependent behavior stems from people's desire to conform with socially-determined \emph{identity} norms.

\subsection{Community Enforcement of Discrimination}
\label{sec:norm1}
\cite{peskiszentes2013} study a dynamic economy where a continuum of agents repeatedly and randomly match in pairs, and each pair has the opportunity to form a short-term profitable partnership. 
Each agent has an unchanging characteristic, their ``physical identity,'' which can be $A$ or $B$. Before agents decides whether to partner up or not, they observe the physical color of their potential partner, and also an additional piece of information, which conveys information about the past partners of this potential match. This additional information is a binary signal, referred to as the agent's ``social identity,'' and it can also be either $A$ or $B$. Unlike their physical identity, an agent's social identity is not fixed, and can change on the path of play (or more importantly, off the path of play). Specifically, if an agent enters a partnership, then their social identity may switch to either the physical or the social identity of their partner.

In this setup, if forming a partnership were a one-time decision, all agents would always choose to enter these relationships. However, in the dynamic context, their desire to form partnerships may be affected by how the agents expect their partnership history to influence  future partnership opportunities. Because an agents' social identity is a signal of their partnership history, this information may be conveyed to future matches, who may react (positively or negatively) to it.
The main result of \cite{peskiszentes2013} is that, under some conditions, there exist equilibria that involve discrimination, in which agents refuse to form relations with potential partners of social or physical identity that does not match their own.\footnote{There are three possible types of discriminatory equilibria. The first involves full segregation, and members of both identities symmetrically discriminate against each other. The other two types are asymmetric discriminatory equilibria, where one identity strongly discriminates against the other, while members of this disfavored identity at most weakly discriminate.} For instance, an identity-$A$ agent may refuse to form a partnership with  an identity-$B$ agent for fear that, if they did so, future $A$ potential partners will refuse them partnership. Indeed, these fears may be well-founded in equilibrium, if the ``equilibrium social norm'' is for $A$ agents to refuse partnership with anyone of either physical or social $B$ identity (remember that partnering up with an identity-$B$ agent may turn someone's social identity from $A$ to $B$). The technology of the changing social identity makes it possible for society to punish not only those who fail to discriminate, but also those who fail to punish non-discriminators.


It is worth noting once more that, in this context, discriminatory equilibria are \emph{not} supported by differences (or perceived differences) in the underlying ability of members of different color-groups. In fact, from a payoff perspective, all workers are identical, both ex-ante and ex-post. Instead, agents act discriminatorily because it is \emph{the social norm} to discriminate; and the social norm is upheld by agents' fears of being punished for not conforming. \cite{peskiszentes2013} argue that this discriminatory mechanism is more than a theoretical possibility in the following quote: ``Are there social institutions with similar features? The Indian caste system, for example, prescribes several rules which prohibit certain kinds of relationships between members of different castes (see \cite{pruthi2004}). These rules are often enforced using the idea of \emph{pollution}. Some castes are considered inherently polluted. A person who accepts a favour or food from a polluted person becomes polluted himself. That is, pollution is treated as something contagious which can only be cured by performing costly rituals.''

\cite{eeckhout2006} also proposes a model in which discrimination and segregation are social norms that arise, despite agents being homogeneous in their payoff-relevant characteristics. He studies a dynamic market where agents bilaterally and randomly match and have the opportunity to form ``marriages,'' which are potentially long term partnerships. Upon meeting, two agents play a partnership game modeled as a potentially repeated Prisoner’s Dilemma. After any transaction, each partner can choose either to remain matched or to terminate the partnership and randomly match with a new agent.

In each period within a partnership, agents weigh a trade-off between defecting, which is myopically valuable, and their cooperative continuation payoff. An agent can always defect and then go back to the matching market in search for a new partnership. This mechanism makes full cooperation, which is efficient, impossible to attain in equilibrium. \cite{eeckhout2006} shows that equilibria often involve incubation strategies, where the norm is for partnerships to be started ``cautiously,'' with an initial phase of defection, followed by cooperation. This initial trial period makes equilibrium deviation costly, because if an agent returns to the matching market, then they will need to start a new relationship and go through the ``caution'' period once more. This initial costly phase is the deviation punishment that allows cooperation to be sustained in later phases of a long-term partnership.\footnote{\cite{eeckhout2006} is not the first paper to observe that ``trial phases'' in long-term relationships help sustain later cooperative phases. The idea that gradual trust-building has a beneficial effect on discipling long-term cooperation was already present in \cite{datta1996}, \cite{ghoshray1996},  \cite{kranton1996}, and \cite{watson1999}. The main contribution in \cite{eeckhout2006} lays in showing that, beyond the trust-building phase, discriminatory norms can also help support cooperation.}

Beyond the ``trial phase,'' \cite{eeckhout2006} shows that discriminatory norms can also help support cooperation and, as such, may be welfare enhancing.
If agents coordinate on not forming partnerships with members of a different color-group, then mixed matches lead to no cooperative value. But, due to random matching, these mixed matches still occur in equilibrium, and so segregation decreases the deviation value of cheating on a current partner and returning to the matching market. As such, the within-color matches can become more cooperative. \cite{eeckhout2006} shows that such segregating equilibria can attain higher welfare than corresponding color-blind equilibria.

\cite{choy2018} is a more recent paper that also proposes a socially enforced reputation model of group-segregation. As in \cite{eeckhout2006}, segregation acts in equilibrium as a coordinating device that is welfare enhancing. However, the equilibrium structure  proposed in \cite{choy2018} structure features a series of hierarchically ranked groups, with higher ranking groups refusing to interact with lower ranking groups but not vice versa. Both \cite{eeckhout2006} and \cite{choy2018} note that the coordination based on individual identities could be mimicked by other public randomization devices that are not related to social identity. As a retort, \cite{eeckhout2006} remarks: ``...while exogenous public randomization devices may be common, for example, in the case of traffic lights, they are far less common in other environments with decentralized social interaction. Here, the point is precisely that a randomization device is being used and that the one used is readily available from the composition of the population.''

\cite{bramoullegoyal2016} also study a repeated partnership-formation game in which agents belong to different identity groups. In their model, a principal always wishes to form partnerships with high-quality agents (experts), and have no inherent preferences for forming partnerships with members of their own identity. \cite{bramoullegoyal2016} study the circumstances under which it may be beneficial for an identity-group as a whole to only form within-group partnerships, even if that means passing on experts belonging to another group. They call this in-group \emph{favoritism}, and argue that favoritism is a mechanism for surplus diversion away from the society at large and toward the group. They show that, depending on economic frictions in the game, it may be beneficial for groups to favor their own members, even if that is detrimental for the economy as a whole.

Note that in both  \cite{peskiszentes2013} and in \cite{eeckhout2006}, discrimination and segregation is supported by equilibrium punishment strategies according which individuals perceive that the cost of ``cheating'' or ``not collaborate'' with others depends on their identities. Similarly, \cite{harbaughto2014} propose that \emph{opportunistic discrimination} arises when firms perceive that they are less harshly punished for opportunistic behavior against individuals belonging to a minority identity, compared to individuals belonging to a majority identity. In their model, a firm repeatedly interacts with individuals belonging to a population of (majority identified and minority identified) agents. In each stage interaction, the firm may choose to cheat or not cheat on the individual with whom they are interacting. \cite{harbaughto2014} characterize parameter regions under which discriminatory equilibria exist in which the firm's opportunistic behavior against minority (majority) agents is only punished in future interactions with other minority (majority) agents. Consequently, the firm is willing to cheat on minority agents precisely because there are less of them, and therefore as a group they impose punishment only on occasional interactions. Conversely, the firm is not willing to cheat on majority agents, as the company foresees frequent future interactions with other members of that group (and therefore frequent punishment). The prediction that smaller groups --- precisely by virtue of being smaller --- are more susceptible to discrimination distinguishes the model in \cite{harbaughto2014} from other work in this section.


\subsection{Evolutionarily Stable Social Norms}
\label{sec:stable}
The literature on \emph{evolutionary game theory} --- \cite{fosteryoung1990} and \cite{young1993} are early contributions --- sees social norms as self-reinforcing patterns of behavior which emerge spontaneously from the decentralized interactions of many individuals accumulating over time into a set of social expectations. Models in this literature normally pose that agents in a population match with each other over time to repeatedly play some game. When called to play, an agent forms expectations about how their opponent will play based on their own previous interactions (about which they have limited memory). With high probability, each agent then chooses their optimal action based on this prior knowledge; and with some small probability, agents randomize their actions --- this random component is likened to natural variation in other evolutionary processes. Typical analyses then use stochastic dynamical systems theory to compute the distribution of long run  \emph{evolutionarily stable} behavior.

\cite{axtellepsteinyoung2001} study an evolutionary model of bargaining in which the interacting agents may be labelled with different identity tags.\footnote{See also \cite{weisbuch2018} which revisits \cite{axtellepsteinyoung2001}  with a more elaborate model of agent cognition.} These identities do not affect their payoffs in the stage game, but are remembered by agents when they choose their actions to best responses to interactions they've had in the past. In this bargaining environment, the authors show that long-run evolutionarily stable behavior involves an equity norm, in which property is shared equally among claimants, and there are no ``class'' distinctions based on individuals' identities. However, they argue that ``metastable'' norms may comprise discriminatory and inequitable regimes. Under such discriminatory norms,  claimants get different amounts based on observable characteristics that have become socially salient (but are fundamentally irrelevant). Computationally, the authors estimate the time it takes to exit from these discriminatory regimes as a function of the number of agents, the length of agents' memory, and the level of background noise. And indeed, they show that the waiting time increases exponentially in memory length and the number of agents, and can be immense even for relatively modest values of these parameters.

\subsection{Identity as a Social Norm or a Social Asset}
\label{sec:norm2}

In ``Economics and Identity'' (QJE, 2000), Akerlof and  Kranton consider how identity, a person's sense of self, affects economic outcomes. They consider an economic environment where each person starts out with a ``social identity'' -- for example, their gender. They posit that each social identity is associated with a class of \emph{prescribed behaviors}, which specify how members of each social identity group are expected to act. In choosing how to behave, each member of a society cares not only about some direct payoff they get from a behavior, but also about whether that behavior ``conforms'' with the prescription for their own identities. Specifically, every agent dislikes (at least to some extent) acting in ways that do not conform with behaviors prescribed to their own underlying identity.
Naturally, in a model where agents have an underlying preference for conforming with their (gender) identity, two agents who differ only in their identity-memberships may choose to behave differently. \cite{akerlofkranton2000} observe, amongst other applications, that gender discrimination in the workplace, may be attributed to people's desire to conform with their gender identities.\footnote{The fact that people's labor supply choices are affected by their desire to conform to their social identities is empirically well documented. For example, \cite{oh2023}) documents in a field experiment in India that ``workers are less willing to accept offers that are linked to castes other than their own, especially when those castes rank lower in the social hierarchy.''} They observe that many jobs are socially gendered; they entail tasks that seen as either ``appropriate for men'' or ``appropriate for women,'' and posit that women will dominate jobs whose requirements match construed female attributes, while men eschew them (and vice-versa). The desire to belong or conform to identity norms has also been used in the literature as a driver of many other phenomena; for example, ethnic and racial conflicts, as in \cite{sen2007}.

More recent developments in this literature propose models of the evolution of \emph{identities as social norms}.  These are evolutionary models, where, in the short run, agents take norms and social categories as given and choose their behavior in order to maximize their utility (including their identity-related utility). But in the long term these (myopically) chosen actions themselves determine the evolution of the identity norms and social categories. 
For example, \cite{carvalhopradelski2022} model the evolution of identity-specific norms based on groups' representation in different activities. Similarly, \cite{akerlofrayo2020} assume that activities are more identity-appropriate when more members of the identity group engage in them. Using a different approach, \cite{akerlof2017} proposes a model of ``identity-formation'' where agents choose how much effort to dedicate to two different activities, as well as their values over the two activities. In social interactions, agents derive value from self-esteem -- how much they excel at activities they value -- and peer-esteem -- how much they excel at activities valued by their chosen peers. \cite{akerlof2017} applies this model to study people's choices to conform or differentiate from their peer groups.

\cite{mailathpostlewaite2006} introduce the notion of identity as a \emph{social asset}, an attribute that has value only because of the social institutions governing society.
\cite{mailathpostlewaite2006} study a matching model, where men and women pair up, consume, and have children. Each person cares about their own consumption, as well as their future kid's consumption. People differ in terms of their wealth -- which is inherently valuable because couples consume jointly -- as well as in terms of a heritable identity attribute that is independent of income and does not directly enter people's utility functions (say, blue eyes). 
There are equilibria where this payoff-irrelevant attribute is ignored, and people match only based on their wealth. 
However, suppose that in this society people with blue eyes are considered more desirable mates -- that is, people are willing to trade a high-wealth mate for a slightly less wealthy one, but with blue eyes. In that case, people would prefer to have kids with blue eyes, because they will be more successful in the matching market, when their time comes. But if people prefer to have kids with blue eyes, and blue eyes are a heritable attribute, then they necessarily prefer to find a partner with blue eyes. Consequently, people's preferences for blue eyes may be self-fulfilling. In that case, \cite{mailathpostlewaite2006} say that blue eyes are a \emph{socially valuable} identity asset, despite them not being intrinsically desirable.

\section{Discriminatory Institutional Design}
\label{sec:design}


In this section, we take the perspective of a mechanism designer (a manager, or a regulator, for example), who wishes to set rules that guide people's behavior in an institution with the goal of maximizing some objective, be it some notion of welfare, efficiency, or profit. 
The papers reviewed here show that, in different contexts, asymmetric mechanisms can be optimal because they improve the overall incentives when agents have hidden actions -- for example, choose to engage in criminal activity or exert effort at their job. They also demonstrate that asymmetric mechanisms can be used to coordinate the actions of multiple agents within an organization. This perspective introduces the idea that discrimination in institutions can exist \emph{by design} to serve the purpose ascribed to the designer.

\subsection{Discriminatory Monitoring}\label{sec:design1} \cite{eeckhoutpersicotodd2010} study a problem of crime deterrence, and argue that treating observably equal agents asymmetrically, with different police-monitoring intensities, can increase the police's effectiveness at deterring crime. To explain the mechanics of this result, they introduce an example: ``Consider a population of 100 citizens, half of whom would never commit a crime, and half of whom would commit a crime unless they are certain that they will be caught. A citizen's propensity to commit a crime is unobservable to the police. The police resources are such that they can check only 50 citizens. Suppose that the police check citizens at random (note that all citizens look the same to police), so that each citizen has a probability $1/2$ of being checked.
Then, only the high propensity citizens will commit a crime, giving rise to a crime rate of $1/2$. Suppose now that half of the citizens have blue eyes, half have brown eyes and that eye color is known to be independent of the propensity to commit a crime. Nevertheless, suppose that police crack down on brown eyed citizens and check them all and completely ignore the blue eyed citizens. Then no brown eyed ever commits a crime because they are sure that they would be caught, and only those blue eyed citizens commit a crime who have high criminal propensity. Thus, the
crime rate with a crackdown on brown eyed persons is $1/4$, which is lower than the crime rate of $1/2$ obtained without crackdowns.''

The example shows that, given a resource constraint that requires the police to monitor only half the citizens, a higher deterrence rate can be achieved by (committing to) concentrating all the police resources on only half of the citizens, and letting the other half be free of any monitoring. This result would be more trivial if the police were able to concentrate their efforts only on citizens they know to be more likely to commit crime, but the example in \cite{eeckhoutpersicotodd2010} shows that, even if people's propensity to commit crime is unobservable to the police, ``crackdown'' deterrence can be more effective. To study the problem of designing crackdowns more generally, \cite{eeckhoutpersicotodd2010} introduce a more general model in which a police force with limited capacity monitors a population made up of individuals with heterogeneous and unobservable criminal propensities. Their main result characterizes optimal monitoring as a function of the distribution of criminal propensities, showing that this policy often involves dividing the population in two groups  which are monitored with different intensities.\footnote{\cite{persico2002} studies a version of the monitoring problem without assuming that the police can commit to a monitoring strategy and shows instead that, even if two subgroups of the population have different propensities to commit crime, it can be efficient to force the police to monitor both groups with the same intensity. Even without differential policing, \cite{verdierzenou} argue that crime rates in populations with the same criminal propensities can arise as the result of a ``self-fulfilling prophecy'' in which agents with different population memberships segregate in more central/less central neighborhoods.}



.



\subsection{Asymmetric Contests}\label{sec:design2}
In the deterrence context, differential monitoring can help reduce total crime by reallocating incentives to commit crimes between groups. Similarly, there are other contexts in which incentives can be reallocated between groups in a manner that improves the principal's outcome of interest. For example, many economic prizes, such as promotions or bonuses, are allocated via contests. In a seminal contribution, \cite{meyer1991} considers the problem of a contest designer who wishes to allocate a promotion to one of two workers. The designer observes the workers' relative outcomes in multiple rounds of a contest, where their outcomes are positively related to their underlying abilities -- the designer is boundedly rational, in that they can only observe the workers' ranks. 

The main question \cite{meyer1991} asks is whether the designer would benefit, in terms of improving the probability of promoting the better worker, from biasing the contest (say, by giving a ``head start'' to one of the two workers). She first shows that the designer always benefits from biasing later rounds of the contest in favor of the winner of earlier rounds, reinforcing the likely ability advantage of the ``leader.'' At that point, the beneficial ``discrimination'' is simply an exacerbation of already existing differences between the two workers, rather than an unequal treatment of two equal workers. However, Meyer (1991) also shows some conditions under which the designer would like to bias the contest even at the initial round, where the two workers are still symmetric from the designer's perspective. 

\cite{kawamurabarreda2014}show that in contests with strategic agents -- workers are non-strategic in \cite{meyer1991} -- it is also in the contest-designer's interest to create bias, improving one agent's success probabilities at the expense of others'. \cite{drugovryvkin2017} show that the designer often benefits from biasing contests even if their objective is not to improve the probability of promoting the better worker, but rather to maximize aggregate effort, or the winner's effort. \cite{dengfangfuwu2023} highlight information disclosure as a separately valuable tool in biasing optimal contests. They show that, if the principal wishes to maximize the contest winner's effort, then the optimal contest is biased in two separate, and opposing, ways: it preferentially informs one of the competitors about the value of the contest's prize, and favors the other competitor in terms of the contest's scoring rule.
\cite{mealemnitzan2016} survey the literature on biased contests, and relate the optimality of asymmetric contests to the optimality of asymmetric mechanisms in auction contexts.\footnote{Both the auctions and the contests literature also consider a question of whether ``leveling the playing field'' is gainful in contexts where participants are ex-ante asymmetric. The idea is that an uneven playing field may discourage effort (or bids) from underdogs, who understand that their probability of winning is reduced. Leveling the playing field across ex-ante asymmetric players can  increase welfare by inducing greater effort or bids, as shown in the context of auctions with toeholds in \cite{bulowhuangklemperer1999} and in research contests in \cite{chegale2003}. More recently, \cite{hossainmorgan2022} show that a similar channel for welfare improvement is that affirmative action that levels the playing field induces greater rates of participation of disadvantaged players. In the context of contests, \cite{drugovryvkin2022} show that leveling the playing field can either encourage or discourage overall effort, depending on details of the contest environment.}

\subsection{Asymmetric Career Paths} \cite{atheyaveryzemsky2000} study a model where a firm designs career paths for its employees. The firm is an overlapping generations environment, where the ``diversity'' among older-generation workers affects the career prospects of younger-generation workers through \emph{mentoring}. In each period, there is a new population of entry-level workers, characterized by their ability and their group membership. There are two groups, $A$ and $B$, and half of the entry-level workers belongs to each group.\footnote{In a recent paper, \cite{mullerittenory2022} extend \cite{atheyaveryzemsky2000} to account for differential group sizes, examining steady-state outcomes and policy under majority bias.} Additionally, the agents' abilities are equally distributed in the two groups. 
There are also agents in upper-level positions, whose group-memberships are not necessarily evenly split. We say that the \emph{majority group} is the one that has most of the upper-level positions; the other group constitutes the \emph{minority}.

Entry-level employees augment their initial ability by acquiring specific human capital in mentoring interactions with upper-level employees. Importantly, an entry-level employee acquires more human capital from mentoring when the firm has more upper-level employees who match her type.
\cite{atheyaveryzemsky2000} consider the problem of the firm who decides a rule to promote employees from entry-level positions to upper-level positions. Any entry-level employee who is not promoted exits the model, and the entry-level positions are replenished with a new cohort. Similarly, all upper-level employees leave the model at the end of each period, and are substituted by promoted entry-level workers. The firm's promotion decisions may depend on employees' abilities and acquired human capital, as well as their group memberships. 

The firm's optimal promotion rule balances two forces. On the one hand, the firm's myopic optimal decision is to promote the most productive agents, accounting for both their inherent abilities and human capital acquired through mentoring. In this respect, the firm is more likely to promote majority workers, who receive more mentoring. On the other hand, the firm has a forward-looking goal to promote agents so as to achieve a desired level of upper-level diversity in the long run. At least some level of diversity is desirable, because the firm wishes to have minority workers with high inherent ability receive good mentoring. 

The paper has two types of results. First, they study what is the optimal bias the firm should implement in their promotion rule in order to balance their myopic and forward-looking goals. In general, they find that the optimal bias need not favor the minority, even if there are decreasing returns to having more mentors of a given type. They explain: ``Because majority employees are better mentored, their promotion rates can be higher than those of minorities, leading the firm to care more about the effective mentoring of majority than minority employees. As a result, a profit-maximizing firm will bias its promotions to favor increased diversity only if there are sufficiently decreasing returns to mentors of a given type.''

Second, they characterize long-run diversity in the firm, under the optimal promotion rule, as well as features of employees' careers. They show that diversity of the upper level can converge to multiple steady states, which can range from full diversity to complete homogeneity. Moreover, equilibria can exhibit a ``glass ceiling'' phenomenon, where the minority in the upper-level starts increasing, but the progress is stalled by the group-based mentoring dynamics, before full diversity is achieved.

\subsection{Asymmetric Contracts towards Coordination} So far, all the work presented in section \ref{sec:design} centers on the idea that if an organization wants to maximize some aggregate variable like total effort or total welfare of its constituent members by designing contracts, contests, or career paths, it might find it useful to institute asymmetric designs that provide stronger incentives to some groups than to others. \cite{winter2004} also argues that asymmetric contract design may be gainful to an organization, but due to a very different mechanism: discriminatory contracts can be effective at creating \emph{coordination} of efforts across individuals in an organization.

He proposes a model of an organization in which the success of a project depends on the contribution of effort from multiple workers. These workers can only be rewarded based on the overall success or failure of the joint project, so the return each worker receives from their own effort also depends on the efforts made by all other workers, making their effort choices strategic complements to each other. Due to this strategic complementarity, depending on the contract proposed by the organization, there may be multiple equilibria, in some of  which all workers exert effort and in some of which that is not the case. \cite{winter2004} asks the following design question: what is the cheapest contract the organization can design that guarantees that the \emph{only equilibrium} is one in which all workers exert effort?

The paper shows that asymmetric contracts --- that provide different rewards for joint success to workers that are equal --- are effective at coordinating the effort of the multiple agents in the organization and therefore delivering ``full effort'' as the unique equilibrium at the lowest possible cost. To understand, suppose there are two workers whose tasks are complementary in their effect on the probability of success of a joint project; and suppose the firm pays $v>0$ to each of these workers if the joint project succeeds. If $v$ very high, then each worker is willing to put in effort at their own task to increase the probability of success, regardless of the effort decision of their partner. Therefore, if $v$ is very high, there is a unique equilibrium in which both workers exert effort. If instead the value of $v$ is at an intermediary range, each worker may still be willing to put in effort, so long as they believe their partner will do the same. 
In this case, there is an equilibrium in which both workers exert effort; but also an equilibrium in which both workers shirk. The conclusion is that, if the contract is symmetric, it takes a very high $v$ to implement full effort as the unique equilibrium. 

Now suppose the organization designs an asymmetric contract, in which the first worker receives a very high value $V$ if the joint project succeeds, while the second worker receives an intermediary value $v$ after a success. Then the first worker necessarily puts in effort in equilibrium, independently of the effort decision of their partner (because $V$ is very high). The second worker, with the intermediary value $v$, is willing to put in effort only so long as their partner does so. In this case, there is a unique equilibrium in which both workers exert effort: the first one because effort is a dominant strategy; the second one because they understand that their partner will necessarily put in effort, and therefore choose to do so as well. Of course, the unique implementation of effort is achieved in this case at a cheaper cost than in the case of symmetric contracts (one very steep incentive contract and one intermediary incentive contract, as opposed to two very high incentive contracts). In this context, the asymmetry in the contract explores the complementarity between workers' strategies to attain the implementation of full effort at a lower cost. 

The cost-minimizing contracts found in \cite{winter2004} discriminate equal workers into different ranks within an organization, and create coordination by making it clear to every worker what their rank is (their own incentive contract) as well as the ranks and incentive contracts of all of their fellow workers. The fact that  each worker knows the contracts received by all their partners is an assumption the framework of \cite{winter2004}. This assumption is challenged in a later paper, \cite{halaclipnowskirappoport2021}, who argue that firms rarely disclose information on their employee's contractual terms, and also tend to discourage or even prohibit employees from discussing this information with each other. With that in mind, \cite{halaclipnowskirappoport2021} propose a different exercise in the environment studied by \cite{winter2004}, which considers that a cost-minimizing firm can design not only the contract received by each worker, but also the information each worker receives about the contract assigned to their fellow employees. The paper shows that the solution to this broader design problem involves withholding from workers information about their partners' incentive contracts, and is non-discriminatory, in the sense that it provides equal incentive contracts to equal workers. 

A policy-relevant takeaway from the contrast between the solution to the problem in \cite{winter2004} and the solution to that in \cite{halaclipnowskirappoport2021} is a relation between contract transparency and discrimination. Contract transparency is commonly pointed to as a tool to combat pay discrimination in companies; the idea being that workers can more easily demand equal treatment from employers if they are aware of the terms of contracts offered to their fellow employees. This logic has recently motivated regulatory interventions aiming to improve transparency inside firms. The models discussed above suggest instead that mandating contract transparency in firms may have an adverse effect on discrimination: the two results together indicate that a firm’s optimal incentive scheme is discriminatory if and only if contracts are required to be public.

\newpage
\bibliographystyle{aer}
\bibliography{referencesfile}

\end{document}